\documentclass[prl,aps,showpacs,twocolumn,superscriptaddress,amsmath,amssymb,floatfix]{revtex4}
\usepackage{graphicx}
\usepackage{color}

\newcommand{\be}{\begin{equation}}
\newcommand{\ee}{\end{equation}}

\newcommand{\ah}{\hat{a}}
\newcommand{\ahd}{\hat{a}^\dagger}
\newcommand{\ch}{\hat{c}}
\newcommand{\chd}{\hat{c}^\dagger}
\newcommand{\ket}[1]{|#1\rangle}

\begin{document}
\title{Extracavity quantum vacuum radiation from a single qubit}
\author{S. De Liberato}
\affiliation{Laboratoire Mat\'eriaux et Ph\'enom\`enes Quantiques,
Universit\'e Paris Diderot-Paris 7 and CNRS, UMR 7162, 75205 Paris
Cedex 13, France} \affiliation{Laboratoire Pierre Aigrain, \'Ecole
Normale Sup\'erieure, 24 rue Lhomond, 75005 Paris, France}
\author{D. Gerace}
\affiliation{CNISM and Dipartimento di Fisica ``A. Volta'',  Universit\`a di Pavia, I-27100 Pavia, Italy}
\author{I. Carusotto}
\affiliation{BEC-CNR-INFM and Dipartimento di Fisica, Universit\`a
di Trento, I-38050 Povo, Italy}
\author{C. Ciuti}
\email{cristiano.ciuti@univ-paris-diderot.fr}
\affiliation{Laboratoire Mat\'eriaux et Ph\'enom\`enes Quantiques,
Universit\'e Paris Diderot-Paris 7 and CNRS, UMR 7162, 75205 Paris
Cedex 13, France}

\begin{abstract}
We present a theory of the quantum vacuum radiation that is generated by a fast modulation of the vacuum Rabi frequency of a single two-level system strongly coupled to a single cavity mode.
The dissipative dynamics of the Jaynes-Cummings model in the presence of anti-rotating wave terms is described by a generalized master equation including non-Markovian terms.
Peculiar spectral properties and significant extracavity quantum vacuum radiation output are predicted for state-of-the-art circuit cavity quantum electrodynamics systems with superconducting qubits.
\end{abstract}
\pacs{
03.70.+k; 
42.50.Pq, 
85.35.Gv 
}

\maketitle 
Cavity quantum
electrodynamics (CQED) is a very exciting and active research field of
fundamental quantum physics, characterized by an unprecedented
control of light-matter interaction down to the single quantum
level~\cite{Raimond01}. A number of different systems and
a wide range of electromagnetic frequencies are presently being
explored in this context, including Rydberg atoms in superconductor
microwave cavities~\cite{Raimond01}, alkali atoms in high-finesse optical
cavities~\cite{Brennecke07,CQED=2LS}, single quantum
dots in semiconductor optical nanocavities~\cite{Reithmaier,Hennessy07}, superconductor Cooper pair quantum boxes in microwave strip-line resonators~\cite{Chiorescu04,Wallraff04,Schuster07,Bishop09}.

Most of the research in CQED has so far concerned systems whose
properties are slowly varying in time with respect to the inverse
resonance frequency of the cavity mode. Only very recently, experiments
with semiconductor microcavities~\cite{Gunter09} have demonstrated
the possibility of modulating the vacuum Rabi coupling on a time
scale comparable to a single oscillation cycle of the field.
For this novel regime, theoretical studies have anticipated the
possibility of observing a sizeable emission of quantum vacuum
radiation~\cite{Ciuti05} via a process that is closely reminiscent
of the still elusive dynamical Casimir effect~\cite{DCE}:  the
modulation of the Rabi coupling provides a modulation of
the effective optical length of the cavity, and it is analogous to a
rapid displacement of the cavity mirrors.

A recent paper~\cite{Dodonov} has applied this general scheme to a
Jaynes-Cummings (JC) model in the presence of a fast modulation of
the artificial atom resonance frequency. However, as the
theoretical model did not include dissipation, the predictions
were limited to short times and were not able to realistically
describe the system steady state. In particular, no quantitative
estimation of the extracavity radiation intensity was provided.

In the present Letter, we introduce a full quantum theory to describe the non-adiabatic response of a JC model including both the anti-rotating wave terms of the light-matter interaction and a realistic dissipative coupling to the environment.
While the former terms are responsible for the generation of photons out of the non-trivial ground state~\cite{Ciuti05,quattropani}, radiative coupling to the external world is essential to detect the generated photons as emitted radiation.
Our attention will be focussed on the most significant case of harmonic temporal modulation of the vacuum Rabi coupling of superconducting qubits in circuit CQED systems: for fast, yet realistic~\cite{private,Delsing} modulations of the vacuum Rabi coupling, the photon emission turns out to be significant even in the presence of a strong dissipation.
Furthermore, in contrast to other systems that have been proposed in view of observing the dynamical Casimir effect~\cite{Law,DeLiberato07,Carusotto08,PD}, the intrinsic quantum nonlinear properties of the two-level system should allow experimentalists to isolate the vacuum radiation from the parametric amplification of pre-existing thermal photons.

In addition to its importance concerning the observation of the dynamical Casimir effect, the theory
developed here appears of great interest also from the general point of view of the quantum theory of open systems~\cite{B-P}.
As a consequence of the anti-rotating wave terms in the Hamiltonian, the ground state of the system contains a finite number of photons.
In order for the theory not to predict unphysical radiation from these bound, virtual photons~\cite{Ciuti06}, the theoretical model has to explicitly take into account the colored nature of the dissipation bath.
This suggests circuit CQED systems as unique candidates for the study of non-Markovian effects in the dissipative dynamics of open quantum systems.

\begin{figure}[ht]
\begin{center}
\includegraphics[width=1 \columnwidth,angle=0,clip]{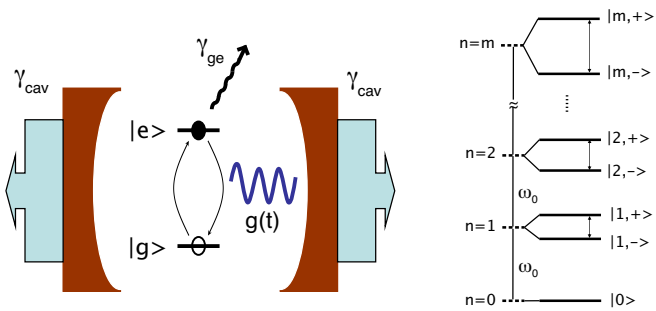}
\end{center}
\caption{Left panel: sketch of the system under consideration. A
single two-level system (qubit) is strongly coupled to a single
cavity mode. A possible realization of such a device consists of a
Cooper pair quantum box embedded in a microwave resonator. Right
panel: schematic representation of the JC ladder of eigenstates of
the isolated system in the absence of modulation, dissipation and
anti-rotating wave terms. In this limit, the eigenstates
$|n,\pm\rangle=(|n,\mathrm{g}\rangle \pm |n-1,\mathrm{e}\rangle)/
\sqrt{2}$ have energies $E_{n,\pm}=n\omega_0 \pm \sqrt{n}g_0$. }
\label{fig_sketch}
\end{figure}

A sketch of the system under consideration is shown in the left panel of Fig.~\ref{fig_sketch}.
The theoretical description of its dynamics is based on the Jaynes-Cummings Hamiltonian:
\begin{multline}
\hat{H}(t)= \hbar \omega_{0} \hat{a}^{\dagger} \hat{a} +
\hbar \omega_{\mathrm{ge}} \hat{c}^{\dagger} \hat{c} +
\hbar g(t) (  \hat{a}  + \hat{a}^{\dagger}) (\hat{c} + \hat{c}^{\dagger})\,.
\label{hamiltonian}
\end{multline}
whose ladder of eigenstates is schematically drawn in the right
panel. Here, $\hat{a}^{\dagger}$ is the bosonic creation operator
of a cavity photon and $\hat{c}^{\dagger}$ is the raising operator
describing the excitation of the two-level system (qubit),
$\hat{c}^{\dagger} \vert \mathrm{g}\rangle = \vert \mathrm{e}
\rangle$, where $\vert \mathrm{g}\rangle$ and  $\vert
\mathrm{e}\rangle$ are its ground and excited states,
respectively; $\omega_0$ is the bare frequency of the cavity mode
and $\omega_{\mathrm{ge}}$ is the qubit transition frequency. The
term proportional to $g(t)$ describes the vacuum Rabi coupling
between the two-level system and the cavity mode and fully
includes those anti-resonant, non-rotating wave processes that are
generally neglected in the so-called rotating-wave approximation
(RWA).

While the RWA has provided an accurate description of most
physical CQED
systems~\cite{Raimond01,Brennecke07,CQED=2LS,Reithmaier,Hennessy07,Chiorescu04,Wallraff04,Schuster07,Bishop09},
it becomes inaccurate as soon as one enters the so-called
ultrastrong coupling regime, i.e. when the Rabi coupling, $g$, is
comparable to the resonance frequencies, $\omega_0$ and
$\omega_{\mathrm{ge}}$. This regime has been recently achieved in
a solid state device consisting of a dense two-dimensional
electron gas with an intersubband transition coupled to a
microcavity photon mode~\cite{Anappara_Rapid}. Values of the
$g/\omega_{\mathrm{ge}}$ ratio of the order of 0.01 (approaching
the so-called fine structure constant limit) have been recently
observed also in circuit CQED systems, and even larger values have
been predicted for more recent unconventional coupling
configurations~\cite{Devoret_alpha}. Fully taking into account the
anti-RWA terms is even more crucial in the experimentally
novel~\cite{Gunter09,private} regime where the Rabi frequency
$g(t)$ is modulated in time at frequencies comparable or higher
than the qubit transition frequency. In fact, in this regime of
non-adiabatic modulation the anti-RWA terms may lead to the
emission of quantum vacuum radiation, a phenomenon that would be
completely overlooked if these terms were neglected. As we show in
this Letter, a significant amount of quantum vacuum radiation with
peculiar spectral features can be already expected for moderate
values of $g/\omega_{\mathrm{ge}}$, i.~e. compatible with already
existing circuit CQED samples.

In order to fully describe the quantum dynamics of the system, the JC model has to be coupled to its environment. A simple description involves two thermal baths, corresponding to the radiative
and non-radiative dissipation channels.
The non-Markovian nature of the baths is taken into account by means of a so-called second order time-convolutionless projection operator method~\cite{B-P}, which gives a master equation of the general form
\begin{equation}
\label{master}
\frac{d\rho}{dt}=\frac{1}{i\hbar}\lbrack \hat{H},\rho
\rbrack+\sum_{j=\mathrm{cav,ge}} \left( \hat{U}_{j}\rho \hat{S}_{j}+\hat{S}_{j}\rho
\hat{U}_{j}^{\dagger}-\hat{S}_{j}\hat{U}_{j}\rho-\rho \hat{U}_{j}^{\dagger}
\hat{S}_{j} \right) \, ,
\end{equation}
where $\hat{S}_{\mathrm{cav}}=(\ah+\ahd)/\hbar$,
$\hat{S}_{\mathrm{ge}}=(\ch+\chd)/\hbar$, and $\hat{U}_j$  are given by
integral operators as
\begin{eqnarray}
\hat{U}_j &=& \int_0^{\infty} v_j(\tau) e^{-i\hat{H}\tau} \hat{S}_j e^{i\hat{H}\tau} \mathrm{d} \tau, \label{Uj}\\
v_j(\tau)&=&\int_{-\infty}^{\infty}
\frac{\gamma_j(\omega)}{2\pi}\lbrack
n_j(\omega)e^{i\omega\tau}+(n_j(\omega)+1)e^{-i\omega\tau}\rbrack
\mathrm{d} \omega.
\end{eqnarray}
The energy-dependent loss rates, $\gamma_j(\omega)$, for the cavity (i.e. $j=\mathrm{cav}$)
and for the qubit transition ($j=\mathrm{ge}$) are related to the density of states at energy
$\hbar \omega$ in the baths, and thus they must be set to zero for $\omega<0$.
In the numerical simulations,
we used the simple form $\gamma_j(\omega)=\gamma_j \Theta(\omega)$ for the non-white
loss rates, where $\Theta(\omega)$ is the Heaviside step function.
In the following, the background number of thermal excitations at energy $\hbar \omega$ in the
 corresponding bath will be set as $n_j(\omega)=0$.
The {\it usually} employed master equation (see, e.g., Ref. \cite{DodonovPRA}) can be obtained from
Eq.~(\ref{master}) by assuming the baths to be perfectly white, i.e. $\gamma_j(\omega)=\gamma_j$. By doing so, one implicitly introduces unphysical negative-energy radiative photon modes, incorrectly leading to the unphysical emission of light out of the vacuum state even in absence of any modulation~\cite{DodonovPRA}.

\begin{figure}[ht]
\begin{center}
\includegraphics[width=\columnwidth,angle=0,clip]{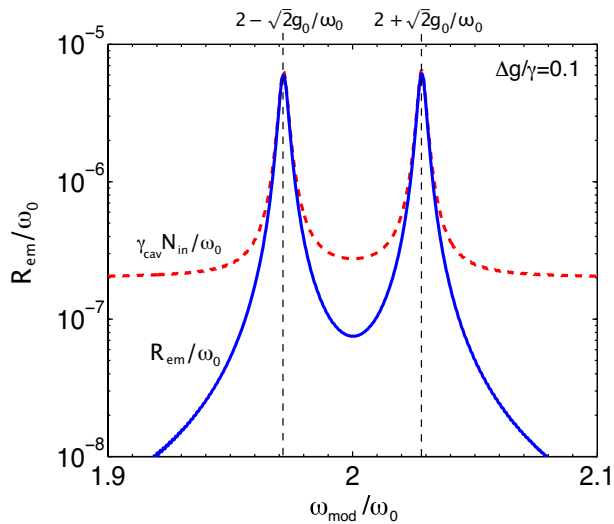}
\end{center}
\caption{Extracavity photon emission rate $R_{\mathrm{em}}$(units
of $\omega_0$) as a function of the modulation frequency,
$\omega_{\mathrm{mod}}$, for a modulation amplitude of the vacuum
Rabi frequency $\Delta g / \gamma =0.1$. Parameters: $\omega_0 =
\omega_{\mathrm{ge}}$;
$\gamma=\gamma_{\mathrm{cav}}=\gamma_{\mathrm{ge}}= 0.002
\omega_0$; $g_0= 0.02 \omega_0$. For comparison, the dashed line
shows the extracavity emission rate
$\gamma_{\mathrm{cav}}N_{\mathrm{in}}$ (where $N_{\mathrm{in}}$ is
the steady-state intracavity photon number) that would be
predicted by the Markovian approximation: note the unphysical
prediction of a finite value of the emission even far from
resonance. \label{fig_extra}}
\end{figure}

In the present Letter, we shall focus on the steady state of the system under a harmonic modulation of the form
\begin{equation}
g(t) = g_0 + \Delta  g\, \sin(\omega_{\mathrm{mod}}\,t) \, \label{modulazione},
\end{equation}
where $\Delta g$ and $\omega_{\mathrm{mod}}$ are the
modulation amplitude and frequency, respectively.
Direct application to the present JC model of the input-output formalism, e.g. discussed
in Ref.~\cite{Ciuti06}, leads to the following expression for the spectral density of extracavity
photons emitted per unit time
\begin{equation}
\mathcal{S}(\omega) =  \frac{\gamma_{\mathrm{cav}}(\omega)}{2\pi} G(\omega)
\end{equation}
in terms of the intra-cavity field spectrum, $G(\omega)$.
As a consequence of the (harmonic) modulation $g(t)$, the spectrum $G(\omega)$ involves a
temporal average over the modulation period $T_{\mathrm{mod}}= 2\pi/ \omega_{\mathrm{mod}}$,
\begin{equation}
G(\omega) = \frac{1}{T_{\mathrm{mod}}} \int_{0}^{T_{\mathrm{mod}}} \mathrm{d} t
\int_{-\infty}^{\infty} \mathrm{d} \tau
 e^{-i\omega \tau} Tr\{
\hat{a}^{\dagger}(t+\tau)\hat{a}(t){\rho}\} \,
\label{spectrally} .
\end{equation}
The cavity field operators $\hat{a}(t)$ are defined here in the Heisenberg picture. The total number of extracavity photons emitted per unit time is given by the spectral integral $R_{\mathrm{em}} = \int_{-\infty}^{\infty} \mathrm{d} \omega \mathcal{S}(\omega)$.
This formula is to be contrasted with the one giving the time-average of the intracavity photon number, $N_{\mathrm{in}}= T_{\mathrm{mod}}^{-1} \int_{0}^{T_{\mathrm{mod}}} \mathrm{d} t \,
Tr\{ \hat{a}^{\dagger}(t)\hat{a}(t){\rho}\} $.

\begin{figure}[ht]
\begin{center}
\includegraphics[width=\columnwidth,angle=0,clip]{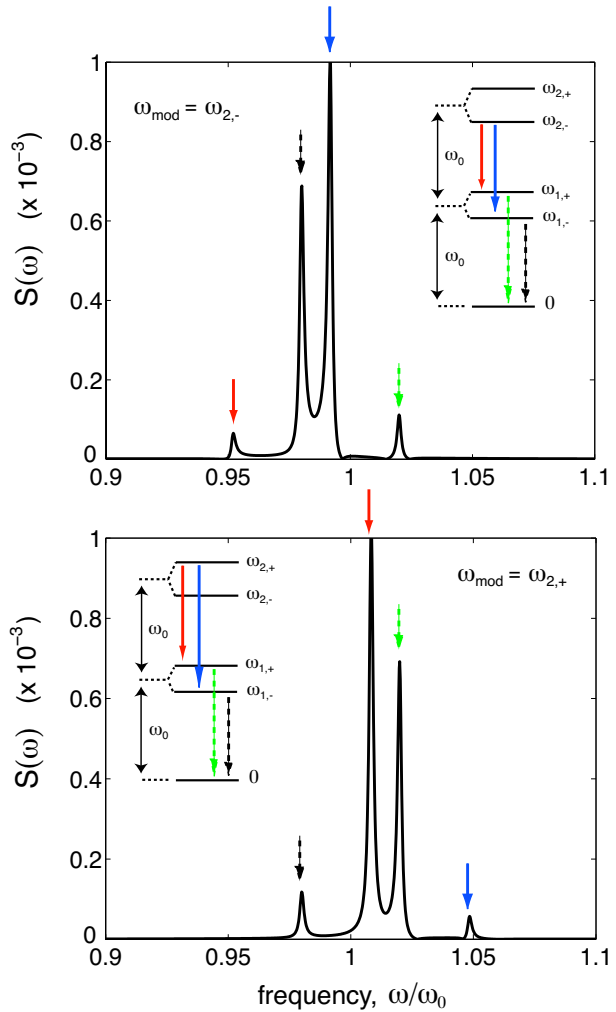}
\end{center}
\caption{Spectrally resolved emission per unit time for given values
of the modulation frequency: $\omega_{\mathrm{mod}}=\omega_{2,-}$ (top panel),
$\omega_{\mathrm{mod}}=\omega_{2,+}$ (bottom panel). The modulation
amplitude is $\Delta g/ \gamma=0.1$. The insets illustrate the
optical transitions responsible for the different
emission lines that are visible in the main panels. For the sake
of clarity, the level spacings in the insets are not in scale.
\label{fig_spectra} }
\end{figure}

The master equation (\ref{master}) is numerically solved by representing $\hat{a}$ and $\hat{c}$
on a basis of Fock number states. The operators $\hat{S}$ and $\hat{U}$ are also numerically built and all the time evolutions are performed by a Runge-Kutta algorithm.
Examples of numerical results are shown in Figs.~\ref{fig_extra}
and \ref{fig_spectra} for the resonant case ($\omega_0 =
\omega_{\mathrm{ge}}$), but we have checked that the qualitative
features do not change when we introduce a finite detuning.
Realistic parameters for circuit CQED systems are considered, as
indicated in the caption.

In Fig.~\ref{fig_extra} we show the steady-state rate of emitted
photons as a function of the modulation frequency,
$\omega_{\mathrm{mod}}$, for the case of a weak
modulation amplitude, $\Delta g/\gamma \ll 1$. In this regime, the
spectra are dominated by two resonant peaks close to
$\omega_{\mathrm{mod}} \simeq \omega_{2,\pm}= 2 \omega_0 \pm
\sqrt{2}\,g_0$ \footnote{For stronger modulation amplitudes (not shown) $\Delta g$, additional, weaker features appear around $\omega_{\rm
mod}\simeq 2\omega_0\pm \sqrt{2/n}\,g_0$  as a consequence of
higher ($n\geq 2$) order processes.}. Thanks to the relatively
small value of $g_0/\omega_0=0.02$ considered here, the position
of the two peaks can be interpreted within the standard RWA in
terms of transitions from the vacuum state to the doubly-excited
states of the JC ladder in the isolated system, $\vert
2,\pm\rangle$; the anti-RWA terms in the Hamiltonian that are
responsible and essential for the quantum vacuum radiation instead 
provide only a minor correction to the spectral position of the
peaks.

As typical of the dynamical Casimir effect, the periodic
modulation of the system parameters is only able to create pairs
of excitations out of the vacuum state. However, in contrast to
the usual case of (almost) non-interacting photons or bosonic
polaritons~\cite{Law,DeLiberato07,Carusotto08}, the nonlinear
saturation of the two-level system  is crucial here to determine the position of the peaks. This remarkable fact provides a unique spectral
signature to separate the vacuum radiation from spurious processes
such as the parametric amplification of thermal radiation. In fact, due to the anharmonicity of the Jaynes-Cummings spectrum, the resonant modulation frequency for the process having the ground state (vacuum) as initial state is different from other processes having a (thermal) excited state as initial state.

The conceptual difference between the emission rate
$R_{\mathrm{em}}$  and in-cavity photon number $N_{\mathrm{in}}$
is illustrated in Fig.~\ref{fig_extra}: standard Markovian
theories would in fact predict the emission rate to be rigorously
proportional to the intra-cavity photon number. Even though a
reasonable agreement is observed around the peaks, this
approximation leads to the unphysical prediction of a finite
emission even in the absence of a modulation $\Delta g=0$ or for a
modulation very far from resonance. Inclusion of the non-Markovian
nature of the baths is able to eliminate this pathology by
correctly distinguishing the virtual, bound photons that exist
even in the ground state from the actual
radiation~\cite{Ciuti05,Ciuti06}.

The emission spectra $\mathcal{S}(\omega)$ at a fixed and resonant value of the modulation frequencies $\omega_{\rm mod}= \omega_{2,\pm}$ are shown in the two panels of Fig.~\ref{fig_spectra}.
Thanks to the relatively weak value $\Delta g/ g_0=0.01$ ($\Delta g/ \gamma=0.1$) of the modulation amplitude considered here, the position of the main emission lines can be again understood in terms of transitions between eigenstates of the JC ladder. As shown in the insets, two spectral lines (red and blue, full lines in the schemes) correspond to radiative decay (emission) of the $\vert 2,\pm\rangle$ states into the lower $\vert 1,\pm\rangle$ states of the JC ladder, while the other two emission peaks (black and green, dashed lines) correspond to the radiative decay of the $\vert 1,\pm\rangle$ states into the ground state. This interpretation is confirmed by the observation that the position of the former (latter) lines depends (does not depend) on the specific value of the modulation frequency $\omega_{\mathrm{mod}}$ chosen. The significant difference of spectral weight between the lines is due to interference effects in the radiative matrix element between JC eigenstates, $\langle 1,\pm\vert \hat{a} \vert 2,\pm\rangle$.
Stronger modulations (not shown) lead to distortion of the spectra as a result of significant spectral shifts and mixing of the dressed states.

The behavior of the exact numerical results can be understood in terms of a simplified two-state model. When the modulation frequency is close to resonance with one of the
$\omega_{\mathrm{mod}} = \omega_{2,\pm}$ peaks, the dynamics of the system is mostly
limited to the $|0\rangle$ and $\vert 2, \pm \rangle$ states, all other states in the JC ladder
being far off-resonant~\cite{CQED=2LS}.
The modulation in Eq. (\ref{modulazione}) is responsible for an effective coupling between such two states, quantified by $\Omega_R\simeq \Delta g/\sqrt{2}$~\cite{Carusotto08}.
As a result, the probability of being in the excited state has the usual saturable Lorentzian shape
\begin{equation}
P_{\ket{2,\pm}} \simeq \frac{(\Delta g)^2/2}{\Gamma^2 + (\Delta
g)^2+4\delta_{2,\pm}^2}\, .
\end{equation}
Here, $\Gamma = [\gamma_{\mathrm{ge}} +
3\,\gamma_{\mathrm{cav}}]/2$ is the total (radiative + non-radiative)decay rate of the
excited $\vert 2, \pm \rangle$ state (in the case $\omega_{\rm cav} = \omega_{\rm ge}$), and
$\delta_{2,\pm}=\omega_{\rm mod}-\omega_{2,\pm}$ is the detuning
of the modulation frequency. By considering all the possible emission cascades (see insets of Fig. 3), the radiative emission rate in the
neighborhood of a peak is then approximately given by
\begin{equation}
R_{\rm em}\simeq P_{\ket{2,\pm}}\,\gamma_{\rm cav}\,\frac{3\gamma_{\rm cav}+2\gamma_{\rm ge}}{\gamma_{\rm cav}+\gamma_{\rm ge}}\, .
\label{emission_anal}
\end{equation}
This analytical expression is in excellent agreement with the exact numerical results for
$\omega_{\mathrm{mod}} \simeq \omega_{2,\pm}$.
It is interesting to note that for typical parameters taken from
state-of-the art circuit CQED devices, such as a resonance
frequency $\nu_{0}=\omega_0 /2\pi\sim 7$ GHz~\cite{Bishop09} and
(overestimated) decay rates $\gamma/2\pi=14$ MHz, a resonant, yet
quite small modulation amplitude $\Delta g/ \gamma =0.1$ can already
lead to a sizeable emission intensity, $R_{\mathrm{em}}\simeq
4 \times 10^4$ photons/second. As clearly shown by the analytical
expression in  Eq.~(\ref{emission_anal}), a further enhancement of the emission
rate can be obtained for much smaller decay rates, such as
$\gamma/2\pi \lesssim  1$ MHz recently measured in the latest
experiments~\cite{Wallraff04,Schuster07,Bishop09}.

In conclusion, we have presented and solved a complete theory of the quantum vacuum emission that is generated from a single mode cavity with an embedded two-level system when the vacuum Rabi frequency of the light-matter interaction is modulated at frequencies comparable to the cavity (emitter)
resonance frequency. Our theory fully takes into account the anti-RWA terms of the light matter-interaction, as well as the radiative and non-radiative dissipation channels. This has required extending the standard master equation treatment to include non-Markovian effects due to the necessarily colored nature of any realistic dissipation bath. The sizable value of the emission intensity that results from our numerical predictions suggests the promise of superconductor Cooper quantum boxes in microwave resonators for studies of quantum vacuum radiation phenomena.

The authors wish to thank M. Devoret, A.V. Dodonov, S. Girvin, and R. Schoelkopf for discussions.

\end{document}